\documentclass[a4paper]{jpconf}

\usepackage{graphicx}
\usepackage{amsmath,amssymb}
\usepackage{cite}
\usepackage{color}

\usepackage{epstopdf}

\def\Jl#1#2{{\it #1} {\bf #2}\ }
\def\CQG#1 {\Jl{Class. Quantum Grav.}{#1}}
\def\GC#1 {\Jl{Grav. Cosmol.}{#1}}
\def\GRG#1 {\Jl{Gen. Rel. Grav.}{#1}}
\def\PRD#1 {\Jl{Phys. Rev. D}{#1}}
\def\PRL#1 {\Jl{Phys. Rev. Lett.}{#1}}

\def\cm{\hspace*{1cm}}

\def\nqq{\hspace*{-2em}}
\def\nhq{\hspace*{-0.5em}}

\def\noi{\noindent}

\def\d{\partial}
\def\al{&\nhq}
\def\lal{&&\nqq {}}

\def\beq{\begin{equation}}
\def\eeq{\end{equation}}
\def\bear{\begin{eqnarray}}
\def\bearr{\begin{eqnarray} \lal}
\def\ear{\end{eqnarray}}
\def\earn{\nonumber \end{eqnarray}}
\def\nn{\nonumber\\ {}}

\def\nnn{\nonumber\\ \lal }

\def\eql{\al =\al}
\def\eqv{\al \equiv \al}

\def\Half{{\dfrac{1}{2}}}
\def\half{{\tfrac{1}{2}}}

\def\eps{\varepsilon}

\def\kappa{\varkappa}
\def\e{{\rm e}}

\def\mn{_{\mu\nu}}
\def\MN{^{\mu\nu}}

\def\mn{_{\mu\nu}}
\def\MN{^{\mu\nu}}

\def\cV{{\cal V}}
\def\eqv{\al \equiv \al}
\def\kappa{\varkappa}
\def\ten#1{\mbox{$\times 10^{#1}$}}

\def\wt{\widetilde}
\def\tg{{\wt g}}
\def\tR{{\wt R}}

\def\S{{\mathbb S}}

\def\M{{\mathbb M}}

\def\od{{\overline d}}

\def\Lamef{\Lambda_{\rm eff}}
\def\sss{\scriptscriptstyle}
\def\mD{m_{\sss D}}

\def\rf{\eqref}

\begin{document}

\title{Cosmology in nonlinear multidimensional gravity and the Casimir effect}

\author{S V Bolokhov$^1$ and K A Bronnikov$^{1,2,3}$}

\address{$^1$ Peoples' Friendship University of Russia (RUDN University), 
               6 Miklukho-Maklaya St., Moscow 117198, Russia}
\address{$^2$ Center for Gravitation and Fundamental Metrology, VNIIMS,
               46 Ozyornaya St., Moscow 119361, Russia}
\address{$^3$ National Research Nuclear University MEPhI
            (Moscow Engineering Physics Institute), Kashirskoe highway 31, Moscow, 115409, Russia}
	
\ead{boloh@rambler.ru, kb20@yandex.ru}        

\begin{abstract}
  We study the possible cosmological models in Kaluza-Klein-type multidimensional 
  gravity with a curvature-nonlinear Lagrangian and a spherical extra space, taking into 
  account the Casimir energy. 
  First, we find a minimum of the effective potential of extra  dimensions, leading to a 
  physically reasonable value of the effective cosmological constant in our 4D space-time. 
  In this model, the huge Casimir energy density is compensated by a fine-tuned 
  contribution of the curvature-nonlinear terms in the original action. 
  Second, we present a viable model with slowly evolving extra dimensions and power-law
  inflation in our space-time. In both models, the results formulated in Einstein and 
  Jordan frames are compared. 
\end{abstract}

\section{Introduction. Basic equations}

  The idea of extra dimensions is now firmly established in theoretical physics in many contexts, 
  such as Kaluza--Klein theories, supergravity, strings and M-theory, brane-world theories. 
  This idea provides a powerful methodological framework for many crucial problems: geometric 
  unification of physical interactions, the hierarchy problem, possible variations of fundamental 
  constants as well as searches for realistic cosmological scenarios including inflationary, 
  string or brane backgrounds, see, e.g., \cite{BR-book, Mel} and references therein.

  One of the important problems in multidimensional models is inclusion of quantum vacuum 
  effects due to the compact topology of extra dimensions (the Casimir effect 
  \cite{Chodos85, Milton, Eliz}). 

  In this paper, in the framework of a Kaluza-Klein-type theory including quadratic curvature 
  invariants, we try to construct viable cosmological models taking into account the Casimir 
  energy of massless fields. For simplicity we restrict ourselves to vacuum models with the 
  geometry $\M^4\times \S^n$, where $\S^n$ is an $n$-dimensional sphere of sufficiently 
  small radius. We thus consider a $(D = 4 + n)$-dimensional manifold with the metric
\beq                                                                \label{ds}
        ds^2 = g\mn dx^\mu dx^\nu - \e^{2\beta(x^\mu)} d\Omega_n^2
\eeq
  where $x^{\mu}$ are the observable four space-time coordinates,
  and $d\Omega_n^2$ is the metric on a unit sphere $\S^n$.
  In this space-time, we consider a curvature-nonlinear theory of gravity with the action
\bear                                                         \label{act1}
           S = \Half \mD^{D-2} \int\sqrt{g_D}\,d^{D}x\,
                  \big(F(R) + c_1 R^{AB}R_{AB} + c_2 R^{ABCD}R_{ABCD} + L_m\big),
\ear
  where capital Latin indices cover all $D$ coordinates, $g_D = |\det(g_{MN})|$,
  $F(R)$ is a smooth function of $D$-dimensional scalar curvature $R$, 
  $c_1,\ c_2$ are constants, $L_m$ is a matter Lagrangian, and $\mD= 1/r_0$ is the 
  $D$-dimensional Planck mass, thus $r_0$ is a fundamental length in this theory.

  We use the system of units with $c = \hbar =1$. As $L_m$, we consider the Casimir 
  energy density in the geometry \rf{ds}. Our goal is to find viable 
  models describing the present-day Universe.  To this end, following the methodology of 
  \cite{BR-06, BR-book,we-16}, we simplify the problem as follows:

\medskip\noi
{\bf (a)}
  Integrate over $\S^n$, reducing all quantities to 4D variables and $\beta(x^\mu)$; thus we have
\bearr                                                      \label{R4}
        R = R_4 + \phi + f_1,  \cm
        f_1 = 2 n \Box \beta + n(n+1) (\d{\beta})^2,
\ear
  where $R_4 = R_4[g\mn]$ is the 4D scalar curvature,
  $\Box = \nabla^\mu \nabla_\mu$ is the 4D d'Alembert operator,
  $(\d{\beta})^2 = g\MN \d_\mu\beta\d_\nu\beta$, and the effective scalar field is equal to the 
  Ricci scalar of $\S^n$:
\beq                                                         \label{phi}
        \phi (x^\mu) = \mD^2\,n (n-1)\,\e^{-2\beta (x^\mu)}.
\eeq
\medskip\noi
{\bf (b)} Suppose slow variations of all quantities as compared with the 
    $D$-dimensional Planck scale, i.e., associate each derivative $\d_{\mu}$ 
    with a small parameter $\eps$ and neglect all quantities of orders higher than $O(\eps^2)$ 
   (see \cite{BR-06, BR-book}). This approximation is justified in almost all thinkable situations.

\medskip\noi
{\bf (c)} The 4D formulation of the theory has the form of 
   a scalar-tensor theory in a Jordan conformal frame. We perform a transition   
   to the Einstein frame, more suitable for studying the dynamics of the 
   scalar field $\phi$ since in this frame it is minimally coupled to the 4D curvature.

\medskip
  In the expression (\ref{R4}), only $\phi$ has the order $O(1)$ while 
  $f_1$ and $R_4$ are $O(\eps^2)$. Therefore,
\bearr                                                       \label{Fapprox}
            F(R) = F(\phi + R_4 + f_1 ) \simeq F(\phi) + F' \cdot(R_4 +f_1 ) + o(\eps^2),
\ear
  where $F' = dF/d\phi$. Thus the 4D (Jordan-frame) action obtained from \rf{act1} 
  takes the form
\bearr                                   \label{act2}
            S = \Half \cV \mD^2 \int \sqrt{g_4}d^4 x \Big[\e^{n\beta} F' R_4 
	  + K_J (\d \beta)^2 - 2 V_J(\phi) \Big],
\ear
  where $\cV(n) = 2\pi^{(n+1)/2}\big/\Gamma(\half(n+1))$  is the volume of a unit sphere 
  $\S^n$, and
\bear
         K_J \eql \e^{n\beta}\big[n(n{-}1)(4\phi F'' - F') 
			+ 4(c_1+c_2)\phi \big],
\nn
	V_J(\phi) \eql \Half \e^{n\beta} [-F(\phi) - c_J \e^{-4\beta} 
				+  2 C_n r_0^{-2} F' \e^{-(n + 4)\beta}].   
\ear
  The dimensionless constants $C_n$ are factors characterizing the Casimir energy 
  density \cite{Candelas},\footnote
	 {This expression for the Casimir energy density is only valid for odd $n$, while 
	   for even $n$ the results are not so confident because of an additional logarithmic
	   divergence \cite{Milton}. We therefore consider only odd $n$.}
  written in such a way that it contributes to the $T^0_0$ component of the total stress-energy 
  tensor of matter in the theory \rf{act2}. The constant $c_J$ is defined as
\beq                                                                               \label{cJ}
           c_J = r_0^{-4}  n(n-1) [(n-1)c_1 + 2c_2]        
\eeq 

  A transition to the Einstein frame is carried out using the conformal mapping
\bearr                                                  \label{trans-g}
                g\mn \mapsto \tg\mn = |f(\phi)| g\mn,
\ \ \
                  f(\phi) =  \e^{n\beta}F_\phi.
\ear
   The resulting action in the Einstein frame has the form 
\bear                                                                                           \label{act3}
     S \eql \Half \cV(n) \mD^2 \int \!\!\sqrt{\tg}\,d^4x\, \Big[\tR_4 
			+ K_{\rm E}(\d\beta)^2 - 2V_{\rm E}(\phi)\Big],  
\\                                                             \label{KE}
        K_{\rm E} \eql 6\phi^2 \Big(\frac{F''}{F'}\Big)^2 \! -2 n \phi \frac{F''}{F'}   
                        + \Half n (n+2) + \frac{4(c_1 + c_2)\phi}{F'},                                      
\\                                                              \label{VE}
          V_{\rm E}(\phi) \eqv  \frac {W(x)}{r_0^2} 
	= \frac{\e^{-n\beta}}{2F'^2} \Big[-F(\phi) - c_J \e^{-4\beta}
                               + 2 C_n r_0^{-2} \cV^{-1} \e^{-(n+4)\beta}\Big],
\ear
  where the tilde marks quantities obtained from or with $\tg\mn$.
  In what follows we will try to build viable cosmologies in the theory \rf{act3} that follows
  from \rf{act1} under the above assumptions. To be consistent with observations in 
  the present-day Universe, we should require:     

\medskip\noi
         1. The space-time is described classically, therefore the size $r = r_0 \e^\beta$ of the 
         extra dimensions should exceed the fundamental length scale $r_0 = 1/\mD$, 
	i.e., $\e^\beta \gg 1$, or $\e^{-\beta} \equiv x \ll 1$.

\medskip\noi
         2. The extra dimensions should not be directly observable, hence 
         $r = r_0 \e^\beta \lesssim 10^{-17}$ cm, which corresponds to the TeV energy scale. 

\medskip\noi
         3. The model parameters should conform to observations: so, our 3D space should  
         expand with acceleration, and the famous Cosmological Constant Problem
         should be somehow addressed.

\medskip
   We will discuss two kinds of models: one based on a minimum of $V_{\rm E}$ and another one
   with slowly evolving extra dimensions. 

\section{A possible stationary state}

  Following \cite{we-16}, let us suppose, for simplicity, $F(R) = -2\Lambda_D + R$, hence  
  $F(\phi) = -2\Lambda_D + \phi$, so that the nonlinearity of the theory is only contained in terms 
  with $c_1$ and $c_2$ in the action \rf{act1}, $F' =1$, $F''=0$, and $f(\phi) =  \e^{n\beta}$.
  Let us also choose for specific calculations $n=3$, then  
\beq                                       \label{Wx}
                W(x) = \lambda x^3 - 3 x^{5} - k_2 x^{7} + k_3 x^{10},
\eeq
  where $x \equiv \e^{-\beta}$; the quantity $W(x)$ is dimensionless as well as the constants
\bearr                \label{VE2}
	\lambda = r_0^2 \Lambda_D, \qquad\ k_1 =  n(n-1)/2, \qquad
	k_2 = r_0^2 c_J/2, \qquad\ k_3 = C_n /\cV.
\ear  
  We seek a minimum of $W(x)$ at some $x=x_0 > 0$,
  corresponding to a stable stationary state of $\beta$ provided that $K_{\rm E} > 0$.
  
\begin{figure}
\centering
\includegraphics[width=7cm,height=4.5cm]{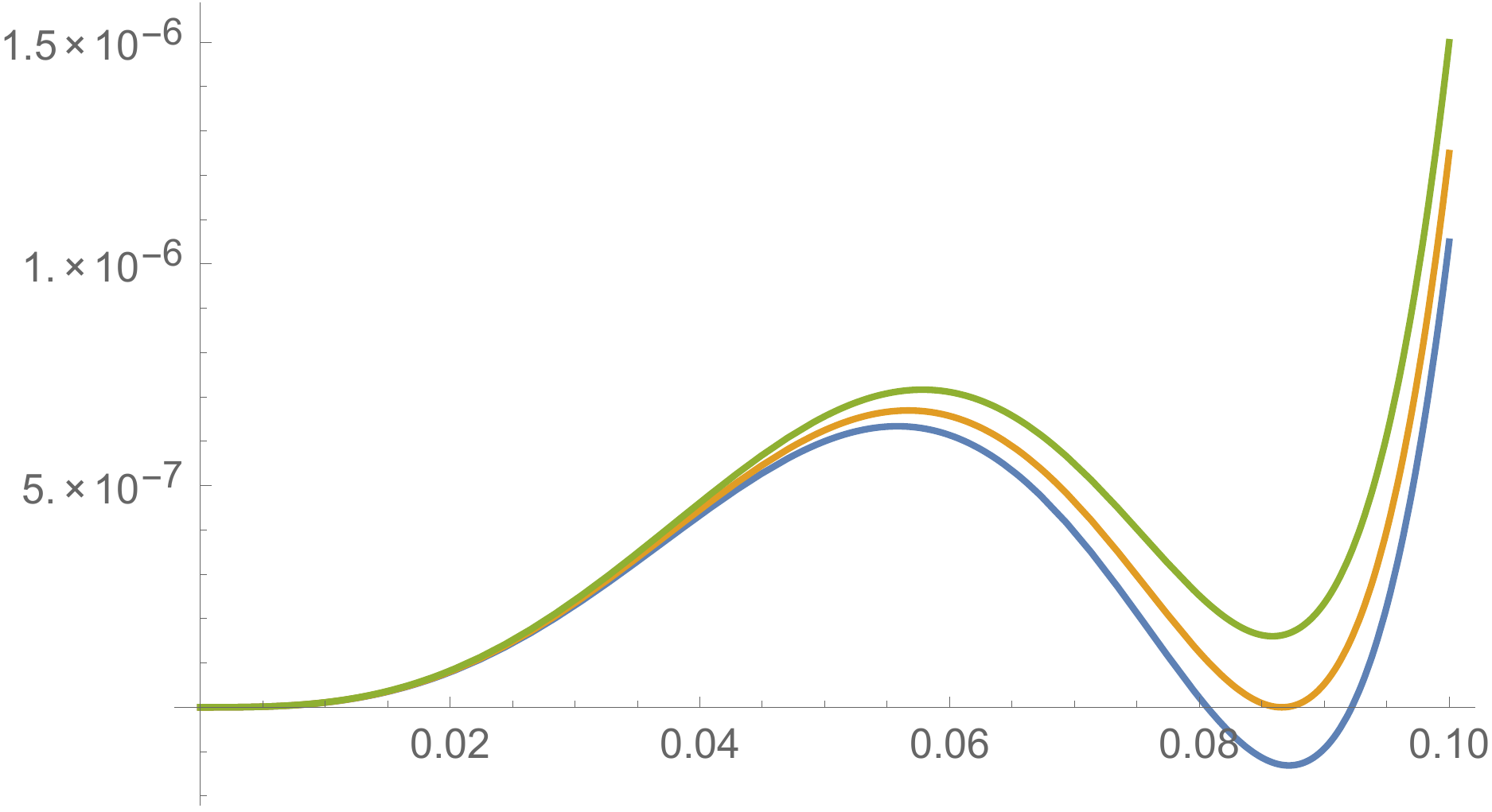}
\qquad
\includegraphics[width=7cm,height=4.5cm]{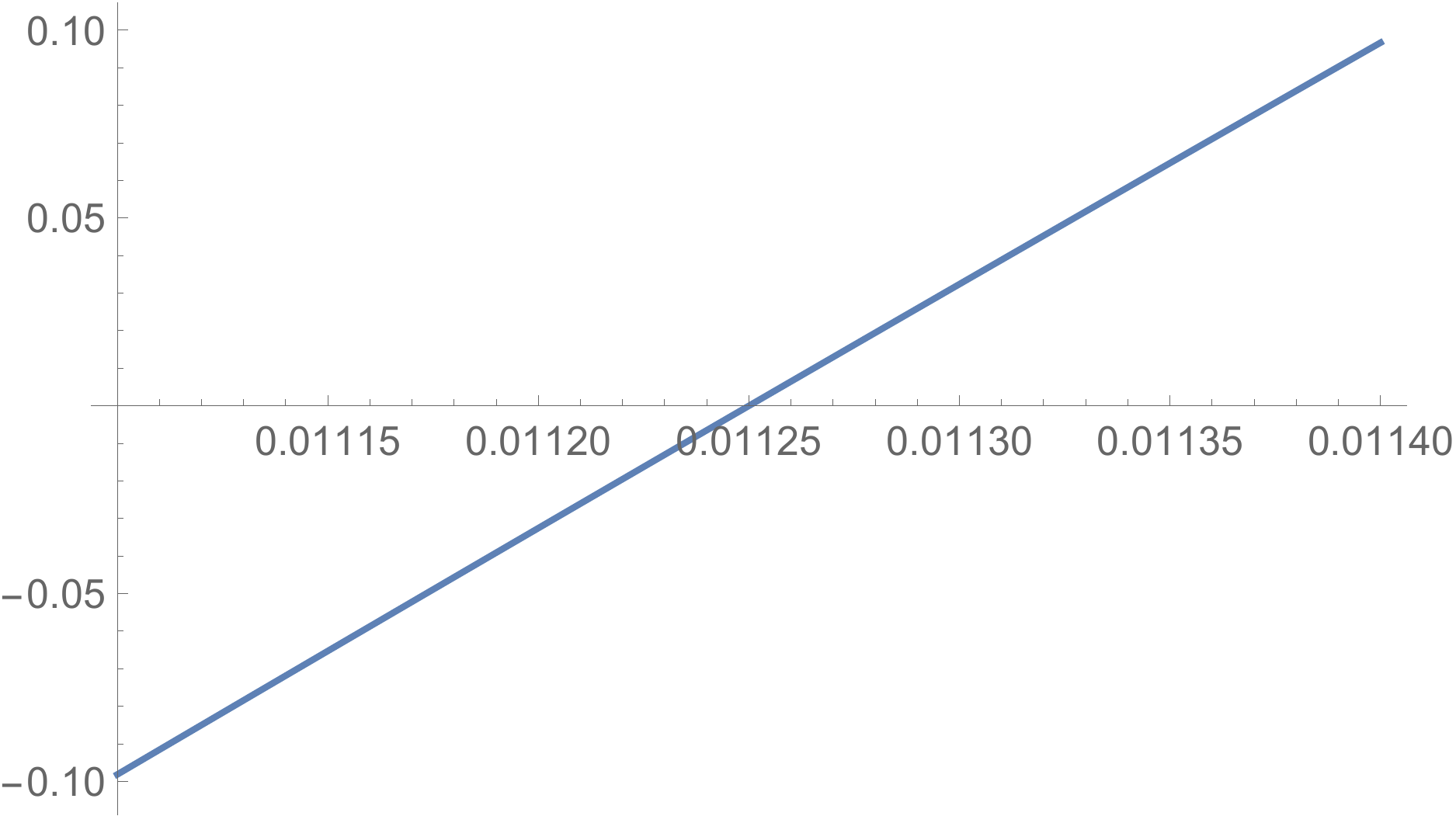}
\caption{\small (a) Plots of $W(x)$ for $n=3$, $F_2=0$, $k_2= -200$, $k_3 = 10^{-4}$, and 
          $\lambda =  0.01105,\ 0.01125,\ 0.0115$ (bottom-up); (b) Plot of $W(x_{\min})$ 
	 as a function of $\lambda$ for the same values of $n,\ F_2,\ k_2,\ k_3$.}  
\end{figure}
  Let us estimate $k_3$ in the Casimir term. According to \cite{Milton}, 
  $C_3 = 7.5687046 \ten{-5}$ for the vacuum density of a single massless scalar field. 
  Furthermore,  $\cV(3) = 2\pi^2$, and one should recall that there is a number of degrees 
  of freedom in 7D space-time, able to increase the Casimir energy density by a factor of 
  $\lesssim 100$. Thus we can take $k_3 \sim 10^{-4}$ for our qualitative estimates. Since the 
  expected values of $x_0$ are small (say, $\sim 0.1$), the term with $k_3$ weakly 
  affects our search for a minimum. Next, without this term but with $k_2 >0$, there 
  cannot be a minimum of $W(x)$ at $x>0$, so we take $k_2 < 0$. Examples of such minima 
  for $k_2 = -200$ are shown in figure 1, where the right panel shows the dependence 
  of $W(x_0)$ on the parameter $\lambda$. One can also verify that $K_{\rm E} > 0$ 
  at $x \lesssim 0.1$. Thus one can easily obtain an arbitrarily small positive 
  value of $W(x_{\min})$ by properly choosing the value of  $\lambda$.

  In cosmology this leads to viable models with $\Lambda >0$, in particular, to the de Sitter
  isotropic model in the absence of matter other than the Casimir vacuum contribution.   

  For further estimates, we should decide which conformal frame corresponds to 
  observations, and this in turn depends on how fermions should be described in a 
  (so far unknown) underlying theory of all interactions. We consider two opportunities, 
  the Einstein frame with the action \rf{act3}, and the Jordan frame, directly obtained 
  from the D-dimensional theory. 

  In the {\bf Einstein frame}, by \rf{act3}, the 4D Planck mass is $m_4 = \sqrt{\cV(n)} \mD$, and
  $r_0 = \sqrt{\cV(n)}/m_4$. Hence for $x_0 \sim 0.1$ 
  the size of extra dimensions is $r(x_0) = r_0/x_0 \sim 10 \sqrt{\cV(n)}\,m_4^{-1}$, 
  close to the Planck length $1/m_4 \approx 8\ten{-33}$ cm as long as $\cV(n)$ is not far from unity. 

  The effective cosmological constant is $\Lamef  = W(x_0)/r_0^2 = W(x_0) m_4^2/\cV(n)$,   
  and to conform to observations which require $\Lamef  \sim 10^{-120}$, fine tuning is necessary:  
  $\lambda$ should be close to the value at which $W(x_0)=0$ (see figure 1, right) with an accuracy 
  close to $10^{-118}$.

  The Casimir contribution to $W(x)$ is very large as compared to 
  $10^{-120}$: e.g., with the parameter values used in the figure, this contribution is
  $k_3 x_0^{10} \approx 1.4\ten{-15}$. This comparatively large value is compensated 
  by fine-tuned values of other parameters of the theory.  
  
  In the {\bf Jordan frame}, the 4D Planck mass is related to $m_4$ by
\beq                                                                           \label{r0J}
	\mD^2 = 1/r_0^2 = m_4^2 \cV(n)^{-1} x_0^n. 
\eeq
  Since $x_0 \ll 1$, $r_0$ is in general a few orders of magnitude larger than the Planck length, 
  which at large enough $n$ may be in tension with the invisibility of extra dimensions.

  The effective cosmological constant in the Jordan frame is obtained if we 
  present the integrand in \rf{act2} as $\sqrt{g_4}\e^{n\beta} F_\phi [R_4 - 2\Lamef  
  + \mbox{kinetic term}]$, which leads to $\Lamef  = F_\phi x_0^{-n} W(x_0)/r_0^2$.
  However, expressing $r_0$ in terms of $m_4$, we arrive again at the 
  expression $\Lamef = W(x_0) m_4^2/\cV(n)$. Thus we need the same fine tuning  
  as in the Einstein frame and have the same estimate of the Casimir 
  contribution to $W(x)$, despite another value of the fundamental length $r_0 = 1/\mD$. 

\section{A model with evolving extra dimensions}
 
  A viable model with slowly evolving extra dimensions can be obtained, by analogy with 
  \cite{bmrs13,we-13}, from the same Einstein-frame action \rf{act3}--\rf{VE}, where now we 
  should focus on the minimum of $W(x)$ at $x=0$ (see figure 1, left panel for illustration) and 
  a slow decrease of  $x(t) \equiv \e^{-\beta(t)} \equiv 1/b(t)$ to zero, so that 
  $b(t) \equiv 1/x \to \infty$, still remaining small enough to conform to observations. 
  Consider 4D isotropic, spatially flat cosmologies with $n$-dimensional spherical extra space:
\beq                                                \label{ds2}
               ds^2 = dt^2 - a^2(t) d{\vec x}{}^2 - b^2(t) d\Omega_n^2.
\eeq
  We now keep $n$ arbitrary. Two independent field equations for the unknowns $a(t)$ 
  and $\beta(t)$ are
\bearr    \label{Fr}
             \frac {3{\dot a}{}^2}{a^2} = K_E {\dot\beta}{}^2 + V_E(\beta),
\\ \lal     \label{be}
              2K_E \biggl({\ddot\beta} + \frac {3{\dot a}}{a}{\dot\beta}\biggr)
		+ \frac {d K_E}{d\beta} {\dot\beta}{}^2 + \frac {d V_E}{d\beta} =0.
\ear
  It turns out that the assumption $F = R- 2\Lambda_D$ now does not lead to a good 
  solution. Instead, following \cite{bmrs13, we-13}, we can choose $F = -2\Lambda_D + R^2$, 
  which leads to the following expressions for $K_E$ and $W = V_E$:
\bearr
             K_E = K_0 = \frac 14 [n^2 -2n +12 + 4(c_1+c_2)],
\nnn
             V_E = \frac 1 {8n^2(n-1)^2}\biggl[2\lambda \e^{(4-n)\beta}
                           - [c_J + n^2 (n-1)^2]\e^{-n\beta} + \frac {2C_n}{\cV(n)}\e^{-2n\beta}\biggr],   
\ear
   where we have put for convenience $r_0 =1$, so that $\lambda = \Lambda_D$.  
   One sees that the required minimum of $V_E$ at $\phi=0$ takes place for $n > 4$, and it is 
   clear that the Casimir term with $C_n \sim 10^{-4}$ again can only weakly affect the solution.
   Neglecting also the second term in $V_E$ (with $\e^{-n\beta}$), it is easy to find a 
   slow-rolling solution, satisfying the conditions $|{\ddot\beta}| \ll 3({\dot a}/a){\dot\beta}$ 
   and $K_E {\dot\beta}{}^2 \ll V_E(\beta)$ \cite{bmrs13, we-13}:
\beq                     \label{ab_t}
        a(t) = a_1 (t + t_1)^p, \qquad b(t) = b_0 \Big(\frac {t+t_1}{t_0+t_1}\Big)^{1/\od},
\eeq
  where 
\beq
           p = \frac{K_0}{\od^2},\quad\  \od = \frac{n-4}{2}, \quad\ 
                 b_0 = \biggl(\frac{\sqrt{\lambda/12}}{H_0 n (n+1)}\biggr)^{1/\od}; 
\eeq
  $a_1$ and $t_1$ are integration constants, $t_0$ is the present age of the Universe,
  and $H_0$ is the Hubble constant: $H_0 = {\dot a}/a$ at $t=t_0$. The solution satisfies
  the slow-rolling conditions if $p \gg 1$ (which in turn requires $c_1+c_2 \gg 1$), and in 
  this case it describes power-law inflation and a very slow increase of the size $b(t)$
  of extra dimensions. 

  The 4D Planck mass is $m_4 = \sqrt{\cV(n)} \mD$, and $r_0 = \sqrt{\cV(n)}/m_4$. 
  If $n$ is not too large (say, $n < 25$), we can suppose $\sqrt{\cV(n)}\approx 1$, so
  we work with approximately the conventional 4D Planck units. From \rf{ab_t} we can 
  estimate the input parameter $\lambda$ (the initial dimensionless cosmological constant) 
  in terms of the present-day size $b_0 = \e^{\beta(t_0)}$ \cite{we-16}:
\beq
                 \lambda \approx \frac 3{16} n^2(n-1)^2 b_0^{n-4} \times 10^{-120}.
\eeq
  To keep the extra dimensions invisible, we should put $1 \ll b_0 < 10^{16}$. It is clear that 
  at some admissible values of $n$ and $b_0$ the quantity $\lambda$ will be of the order of unity, 
  for instance, it happens at $n =13$ and $b_0 = 10^{13}$. Thus the Cosmological Constant 
  Problem can be solved in this framework without fine tuning.

  These estimates have been made in the Einstein frame. As to Jordan's, it turns out 
  \cite{we-13} that the same solution leads to a model incompatible with observations:
  the cosmological expansion is too strongly accelerated (for the effective equation-of-state 
  parameter $w$ we have $-w \gg 1$ while observations imply $w \approx -1$) and leads to 
  a big rip. 

\section{Concluding remarks}

  In curvature-nonlinear multidimensional gravity with spherical extra dimensions, we have 
  presented two simple examples of cosmological models approximately describing the present
  stage of the Universe evolution: one with a suitable minimum of the effective potential,
  corresponding to a stable stationary size $b$ of the extra dimensions, 
  the other in which this size is very slowly growing. The second model differs from the first one 
  in the following features: (i) a larger dimension $n$ is required, (ii) no fine tuning of the initial
  parameters is necessary, (iii) the Casimir contribution to the total energy 
  density is insignificant, and (iv) only the Einstein frame is viable.

  Among possible interesting extensions of this work there is a study of variations of 
  fundamental constants depending on the size of extra dimensions, including the 
  Newtonian gravitational constant $G$ and the electromagnetic coupling constant 
  $\alpha$, a possible application of such models to the 
  early (inflationary) Universe with maybe quantum tunneling between different minima of the 
  potential, and, certainly, a consideration of other geometries of extra dimensions, e.g., 
  in the form of products of spherical, toroidal and/or hyperbolic factor spaces.
 
\section*{Acknowledgments}

  We thank Milena Skvortsova for helpful discussions.
  The work of KB was partly performed within the framework of the Center 
  FRPP supported by MEPhI Academic Excellence Project 
  (contract No. 02.03.21.0005, 27.08.2013).
  This paper was also financially supported by the Ministry of Education and Science of the Russian
  Federation on the program to improve the competitiveness of the RUDN University among the 
  world leading research and education centers in 2016--2020, and by RFBR grant 16-02-00602.



\section*{References}
\begin{thebibliography}{99}

\bibitem{BR-book} 
             Bronnikov K A and Rubin S G 2012 {\it Black Holes, Cosmology 
	     and Extra Dimensions} (Singapore: World Scientific).

\bibitem{Mel}
             Melnikov V N 2016 \GC {22} 80.

\bibitem{Chodos85} 
             Chodos A and Myers E 1985 \PRD {31} 3064 

\bibitem{Milton} 
	   Milton K A 2001 {\it The Casimir Effect: Physical Manifestations of Zero Point Energy} 
            (Singapore: World Scientific)

\bibitem{Eliz}
	    Elizalde E {\it et al} 1994 {\it Zeta Regularization Techniques with Applications} 
              (Singapore: World Scientific).

\bibitem{BR-06} 
            Bronnikov K A and Rubin S G 2006 \PRD {73} 124019 

\bibitem{we-16} 
            Bolokhov S V and  Bronnikov K A 2016 \GC {22} 323

\bibitem{Candelas} 
             Candelas P and Weinberg S 1984 {\it Nucl. Phys.} B {\bf 237} 397
 
\bibitem{bmrs13}
	   Bronnikov K A  {\it et al} 2013 \GRG {45} 2509

\bibitem{we-13} 
            Bronnikov K A and Skvortsova M V 2013 \GC {19} 114

\end{thebibliography}
\end{document}